\def\i{{\rm i}}

\def\e{{\rm e}}
\def\la{\lambda}
\def\q{q}

\def\binom#1#2{\left(\begin{array}{@{}c@{}} #1 \\ #2 \end{array}\right)}
\DeclareFontShape{OML}{cmm}{m}{b}{%
   <-> cmmib10}{}
\DeclareMathAlphabet{\mathbf}{OML}{cmm}{m}{b}
\DeclareSymbolFont{boldletters}{OML}{cmm}{m}{b}
\DeclareMathSymbol{\balpha}{\mathord}{boldletters}{11}
\DeclareMathSymbol{\bbeta}{\mathord}{boldletters}{12}
\DeclareMathSymbol{\bgamma}{\mathord}{boldletters}{13}
\DeclareMathSymbol{\bomega}{\mathord}{boldletters}{33}
\DeclareMathSymbol{\bsigma}{\mathord}{boldletters}{27}
\DeclareMathSymbol{\btau}{\mathord}{boldletters}{28}
\documentstyle[aps]{revtex}
\begin{document}
\title{\large
\bf The XXZ spin chain at $\Delta=- \frac{1}{2}$:\\
Bethe roots, symmetric functions and determinants}
\author{J de Gier, M T Batchelor}
\address{Department of Mathematics, School of Mathematical Sciences,\\
Australian National University, Canberra ACT 0200, Australia}
\author{B Nienhuis, S Mitra}
\address{Instituut voor Theoretische Fysica, Universiteit van
Amsterdam,  1018XE Amsterdam, The Netherlands}
\date{\today}
\maketitle 
\begin{abstract}
A number of conjectures have been given recently concerning the
connection between the antiferromagnetic XXZ spin chain at $\Delta =
- \frac12$ and various symmetry classes of alternating sign matrices.
Here we use the integrability of the XXZ chain to gain further
insight into these developments. In doing so we obtain a
number of new results using Baxter's $Q$ function for the XXZ chain
for periodic, twisted and open boundary conditions.
These include expressions for the elementary symmetric functions
evaluated at the groundstate solution of the Bethe roots.
In this approach Schur functions play a central role and enable us to
derive determinant expressions which appear in certain natural double
products over the Bethe roots. When evaluated these give rise to the
numbers counting different symmetry classes of alternating sign
matrices.
\end{abstract}

\section{Introduction}

The XXZ Heisenberg spin chain is a central and much studied model
in statistical mechanics.
It is arguably the best known model solved by means of the
Bethe wavefunction Ansatz \cite{Bethe}.
The groundstate wavefunction at the particular anisotropy value 
$\Delta=- \frac{1}{2}$
has recently been the source of some surprising observations.
At this value Razumov and Stroganov \cite{RS} observed
the appearance of the numbers $A_n = 1, 2, 7, 42, 429, \ldots$, which
count the number of $n \times n$ alternating sign
matrices \cite{MRR,Bress,Propp}.
Here the length of the chain $L$ is odd ($L=2n+1$)
with periodic boundary conditions imposed.
Alternating sign matrices are matrices whose
elements are either $-1,0$ or $1$ such that the elements along each
row and each column alternate in sign. Furthermore, the entries in 
each row and column add up to $+1$.
The numbers $A_n$ are well known to enumerative combinatorialists,
having appeared in other distinct problems such as the enumeration of
plane partitions and generalizations of determinants. 
The one-to-one correspondence between vertex configurations of
the square lattice ice model with domain wall
boundary conditions \cite{Kor}  and ASM's has been well documented
\cite{Bress,Propp}.
In particular it led to Kuperberg's alternate proof \cite{K1,K} of the
alternating sign matrix conjecture \cite{MRR,Bress}.

The numbers $A_n$ appear in the groundstate wavefunction of
the XXZ Heisenberg chain at $\Delta=- \frac{1}{2}$ in three
ways \cite{RS}:
(i) as the ratio of the largest and smallest components in the
groundstate wavefunction, (ii) in the sum of the components, and (iii)
in the sum of the square of the components.
These observations at $\Delta=- \frac{1}{2}$ have been extended in a 
number of directions.
Two other known cases are:
(i) twisted boundary conditions \cite{ABB} with $L$ even
\cite{BdGN,RS-a}, and (ii) open boundary conditions with appropriate
surface fields (the quantum $U_q[sl(2)]$ invariant chain
\cite{ABB,ABBBQ,Pasq}) \cite{BdGN}. 
For both cases the groundstate wavefunction is complex.
Nevertheless, the sums of the wavefunction components and their
squares are real.
The numbers $A_n$ also appear in the twisted case.
However, the open case sees the appearance of other symmetry classes.
Here appear $A_{\rm V}(2n+1)$, the number of $(2n+1) \times (2n+1)$
vertically symmetric alternating sign matrices when $L=2n$ and
$N_8(2n)$, the number of cyclically symmetric transpose complement
plane partitions, when $L=2n-1$. 
The number $N_8(2n)$ is conjectured to be $A_{\rm VH}(4n+1)/A_{\rm
V}(2n+1)$, where $A_{\rm VH}(4n+1)$ is the number of $(4n+1) \times
(4n+1)$ vertically and horizontally symmetric alternating sign
matrices \cite{Rob}.  
These various numbers also appear in the corresponding O(1) loop model,
for which the groundstate wavefunction is real \cite{BdGN}.
Further developments include the combinatorial
interpretation \cite{RS-b,RS-c,PRdG} of the elements of the O(1) loop
model wavefunction and the relation to a one-dimensional
Temperley-Lieb stochastic process \cite{PRdG}.

The numerous conjectures made to date for the various groundstate
wavefunctions at $\Delta=- \frac{1}{2}$ remain to be proved.
In earlier work, Stroganov and co-workers \cite{FSZa,FSZb,Str} have
found an expression for Baxter's $Q$ function \cite{Baxter}
in each of the above cases.
By definition the zeros of the $Q$ function are the Bethe roots.
However, little if any use has been made of this function.
Here we use the $Q$ function results to obtain closed form expressions
for the values of the elementary symmetric functions with the
groundstate Bethe roots as arguments. 
This approach also involves the appearance of the Schur function
and determinants in a natural way.
Ultimately we are led to conjectures for new determinants whose
values are related to the alternating sign matrix numbers.
These results came from observations on the product of certain
combinations of Bethe roots.
Although some results can be proved along the way, the
evaluation of the determinants, involving the
elementary symmetric functions, remains to be done exactly.

The layout of this paper is as follows.
In Section II we collect some necessary results on the Bethe Ansatz
and symmetric polynomials.
In Section III we give our results for the periodic ($L$ odd),
twisted and reflecting boundary cases.
Some detailed working is given in the Appendix.
The paper concludes with some remarks in Section IV.

\section{PRELIMINARIES}

\subsection{XXZ spin chain and Bethe's Ansatz}

We consider the periodic anisotropic XXZ spin chain.
A spin variable lives on each site of the chain taking either up or
down values.
The interaction between two neigbouring spins is described in terms
of Pauli spin matrices by the well known Hamiltonian 
\begin{equation}
H = - \frac{1}{2} \sum_{j=1}^{L} \left( \sigma_j^x \sigma_{j+1}^x +
\sigma_j^y \sigma_{j+1}^y + \Delta \sigma_j^z \sigma_{j+1}^z \right),
\label{eq:XXZham}
\end{equation}
where the anisotropy $\Delta$ is parametrized by
\begin{equation}
\Delta = -\frac{1}{2}(q+q^{-1}).
\end{equation}
We denote the position of the $i$th down spin along the chain by $x_i$. 
The Hamiltonian (\ref{eq:XXZham}) is diagonalized via the Ansatz \cite{Bethe} 
\begin{equation}
\psi(x_1,\ldots,x_n) = \sum_{\pi} A^{\pi_1\ldots\pi_n}
\prod_{j=1}^n z_{\pi_j}^{x_j} \label{eq:eigvec}
\end{equation}
for the form of its eigenvectors.
The sum over $\pi = (\pi_1,\ldots,\pi_n)$ denotes a sum over all
permutations of the numbers $1,\ldots,n$.
Substituting (\ref{eq:eigvec}) into the eigenvalue equation for $H$ 
one finds the eigenvalues to be given by
\begin{equation}
E = -\frac{1}{2}L\Delta - \sum_{i=1}^n (z_i + z_i^{-1} -2\Delta).
\label{eq:energy}
\end{equation}
The amplitudes $A^{\pi_1 \ldots \pi_n}$ are also expressed in the
variables $z_i$ for which the consistency equations 
\begin{equation}
z_i^L = (-)^{n-1} \prod_{j=1}^n \frac{ 1 -2\Delta z_i +z_i z_j}{ 1
-2\Delta z_j +z_i z_j} \label{eq:BAPBC}
\end{equation}
are derived.
%

It will be convenient to make the change of variables  
\begin{equation}
z = \frac{q-w}{q w-1} 
,\qquad w = \frac{z +q}{q z+1}, \label{eq:varchange}
\end{equation}
for which the Bethe equations (\ref{eq:BAPBC}) take the form 
\begin{equation}
\left( \frac{q-w_i}{q w_i-1}\right)^L + \prod_{j=1}^n
\frac{w_i - q w_j}{q w_i -w_j} =0. \label{eq:BAwPBC}
\end{equation}
Up to a normalisation, the amplitudes are given by 
\begin{equation}
A^{\pi_1 \ldots \pi_n} = q^{n(n-1)/2} \prod_{i < j} \frac{1-2\Delta
z_{\pi_i} + z_{\pi_i} z_{\pi_j}}{z_{\pi_i} -z_{\pi_j}} = \prod_{i < j}
\frac{w_{\pi_i} - q^2 w_{\pi_j}} {w_{\pi_i} -
w_{\pi_j}}. \label{eq:amplitude}
\end{equation}
Note that the eigenfunctions (\ref{eq:eigvec}) are symmetric
polynomials in the variables $z_i$.  
All properties of the XXZ spin chain can therefore be expressed in 
symmetric functions of these variables. 
We will review some of the basic properties of symmetric polynomials 
in the next section.  

{}From (\ref{eq:amplitude}) we see that the amplitudes can be written in
terms of the generalized Vandermonde product 
\begin{equation}
\det{}_\la (w_i^{n-j}) = \det{}_\la (w_j^{i-1}) = \prod_{i<j } (w_i + \la\,
w_j), 
\end{equation}
where we have introduced the $\la$-determinant \cite{RR} which can be defined
recursively via Dodgson's algorithm for evaluating determinants. 
If $X^{(1)} = (x^{(1)}_{ij})$ is an $n\times n$ matrix and $Y^{(1)}$
an $n-1 \times n-1$ matrix with each element equal to $1$, we define
new matrices $X^{(k)}$ and $Y^{(k)}$ recursively by  
\begin{eqnarray}
x^{(k)}_{ij} &=& (x^{(k-1)}_{ij} x^{(k-1)}_{i+1,j+1} + \la\,
x^{(k-1)}_{i+1,j} x^{(k-1)}_{i,j+1})/y^{(k-1)}_{ij}, \quad i,j =
1,\ldots, n-k+1,\\ 
y^{(k)}_{ij} &=& x^{(k-1)}_{i+1,j+1}, \quad i,j=1,\ldots,n-k.
\end{eqnarray}
The number $X^{(n)}$ thus defined is called the $\la$-determinant of
$X^{(1)}$.  
For the special value $\la=-1$ this procedure evaluates the
ordinary determinant $\det X^{(1)}$. 
Just as the determinant can be written as a sum over the set of permutation 
matrices, the $\la$-determinant can be written as a sum over the set of 
alternating sign matrices \cite{RR} 
\begin{equation}
\det{}_\la M = \sum_{A\in {\mathcal A}_n} \la^{{\mathcal I}(A)} (1 +
\la^{-1})^{N(A)} \prod_{i,j=1}^n m_{ij}^{a_{ij}},
\label{eq:qdet}
\end{equation}
where ${\mathcal A}_n$ is the set of $n\times n$ alternating sign
matrices, ${\mathcal I}(A)$ is the inversion number of $A$ and $N(A)$
the number of $-1$'s in $A$ (see e.g., \cite{Bress} for the definiton of
${\mathcal I}$). 
The total number of terms in this sum, or equivalently
the number of $n\times n$ alternating sign matrices is given by
\begin{equation}
A_n = \prod_{j=0}^{n-1} \frac{(3j+1)!} {(n+j)!} = \prod_{1\leq i\leq
j\leq n} \frac{n+i+j-1} {2i+j-1}.
\label{eq:asm}
\end{equation}
We will see later that this and related numbers surprisingly show up
in certain combinations of the variables $z_i$ when evaluated at a
particular solution of the Bethe equations (\ref{eq:BAPBC}) for
$q=\e^{\i \pi/3}$.

\subsection{Symmetric polynomials}

This section is included for the convenience of the reader.
The results collected herein can be found in standard textbooks such 
as \cite{Bress} or \cite{Stanley}.

A partition $\mu$ of $k$ is a non-increasing set of integers
${\mu_1,\ldots,\mu_l}$ such that $\mu_1 + \ldots + \mu_l = k$. 
We denote this by $\mu \vdash k$. Partitions define the shape of a
semistandard tableau, which is a two-dimensional array of integers
with the restriction that these integers are non-decreasing across
each row of length $\mu_j$ and strictly increasing down columns. 
An example of a semistandard tableau of shape $(4,4,3,1,1)$ is  
\[
\begin{array}{cccc}
1 & 1 & 2 & 3\\
2 & 3 & 4 & 4\\
4 & 4 & 5\\
5\\
6
\end{array}
\]
A conjugate partition $\mu'$ or conjugate shape is the shape
obtained by reflecting the semistandard tableau of shape $\mu$ in its
main diagonal. For example, the conjugate partition of $(4,4,3,1,1)$
is $(5,3,3,2)$. The integers in the semistandard tableau may be
interpreted as indices of variables, and thus to every semistandard
tableau is associated a monomial in which the power of
each variable is the number of times its index occurs in the
tableau , e.g. for the example above the monomial is given by
\[
w_1^2 w_2^2 w_3^2 w_4^4 w_5^2 w_6.
\]
Given a tableau $T$ the correspondig monomial is denoted by ${\bf w}^T$.  
In this way one may associate with every tableau of shape $\mu$
a polynomial $s_{\mu}$, called the Schur polynomial, by the definition
\begin{equation}
s_{\mu}(w_1,\ldots,w_n) = \sum_T {\bf w}^T,
\end{equation}
where the sum is over all semistandard tableaux of shape $\mu$ with
entries chosen from $\{1,\ldots,n\}$. 
If $\mu \vdash k$, $s_\mu$ is a polynomial of degree $k$. 
We will see later that the Schur function is a symmetric function.

The monomial symmetric function of degree $k$ is defined by
\begin{equation}
m_{\mu} =  \sum_{\pi}{}' \prod_{j=1}^n w_j^{\pi_j},
\end{equation}
where the sum is over all {\em distinct} permutations
\[ \pi=(\pi_1,\ldots,\pi_n)\quad {\rm of\; the\; numbers} \quad
(\mu_1,\ldots,\mu_k,0,\ldots,0). 
\] 
The elementary symmetric function of degree $k$ in $n$ variables
is defined as the monomial symmetric function that corresponds to the
partition with $k$ 1's,  
\begin{eqnarray}
e_0 &=& 1 \nonumber\\
e_1 &=& w_1 + \ldots + w_n \nonumber\\
e_2 &=& w_1w_2 + w_1w_3 + \ldots w_{n-1}w_n \label{eq:esf}\\
&\vdots& \nonumber\\
e_k &=& \sum_{1 \leq i_1 < i_2 \ldots < i_k \leq n} w_{i_1} \cdots
w_{i_k} \nonumber.
\end{eqnarray} 
Given a partition $\mu$, this definition is extended to
\begin{equation}
e_{\mu} = e_{(\mu_1, \ldots,\mu_l)} = e_{\mu_1} \cdots e_{\mu_l}.
\end{equation}
Finally, the complete symmetric function $h_k$ is defined to be
the sum over all monomial symmetric functions of degree $k$, i.e.,
\begin{equation}
h_k = \sum_{\mu \vdash k} m_{\mu}.
\end{equation}
In a similar way as above this is extended to
\begin{equation}
h_{\mu} = h_{(\mu_1,\ldots,\mu_l)} = h_{\mu_1} \cdots h_{\mu_l}.
\end{equation}

The following facts are well known concerning the various functions
defined above. 
\begin{enumerate}
\item [i)]
The elementary and complete symmetric functions are dual to each other
in the sense
\begin{eqnarray}
e_k &=& \det \left(h_{1-i+j} \right)_{i,j=1}^k, \\
h_k &=& \det \left(e_{1-i+j} \right)_{i,j=1}^k,
\end{eqnarray}
where we put by convention $e_{-k}=h_{-k}=0$ for $k>0$.
Their generating functions are given by
\begin{eqnarray}
\sum_{i=0}^{\infty} e_i(w_1,\ldots,w_n) t^i &=& \prod_{j=1}^n (1+w_j
t), \label{eq:genfune}\\
\sum_{i=0}^{\infty} h_i(w_1,\ldots,w_n) t^i &=& \prod_{j=1}^n
\frac{1}{1-w_j t}. \label{eq:genfunh}
\end{eqnarray}
\item [(ii)]
The Jacobi-Trudi identity.

Let $\mu = (\mu_1,\ldots,\mu_n)$ be a partition into at most $n$ part, then
\begin{equation}
s_{\mu}(w_1,\ldots,w_n) = \det \left( h_{\mu_i+j-i} \right)_{i,j=1}^{n}.
\label{eq:JT}
\end{equation}
\item[iii)] 
The N\"agelsbach-Kostka identity.

Let $\mu'$ be the partition conjugate to $\mu$ and $k$ the number of
parts in $\mu'$, then
\begin{equation} 
s_{\mu} = \det \left(e_{\mu'_i-i+j} \right)_{i,j=1}^k.
\label{eq:NK}
\end{equation}
\item[iv)]
The Schur function is equal to a ratio of Vandermonde determinants,
\begin{equation}
s_{(\mu_1,\ldots,\mu_n)}(w_1,\ldots,w_n) = \frac{\det w_i^{n-j+\mu_j}}
{\det w_i^{n-j}}.
\label{eq:Schurdet}
\end{equation}
Usually (\ref{eq:Schurdet}) is taken as the definition of the
Schur function.
\end{enumerate}

\section{Results for $\q = \e^{\i \pi/3}$}

We now turn to the XXZ spin chain in the following cases: 
(i) periodic boundaries and $L=2n+1$,
(ii) twisted boundaries with twist angle $\pi/3$ and $L=2n$, and 
(iii) reflecting boundaries.
In these cases the XXZ chain has a trivial groundstate energy. 
It is to be understood that in this section we take $q=\e^{\i\pi/3}$.

\subsection{Periodic boundaries}

Consider the function $Q_n(w)$, of which the zeros are the
solutions of the Bethe equations (\ref{eq:BAwPBC}),
\begin{equation}
Q_n(w) = \prod_{i=1}^n (w-w_i) = \sum_{l=0}^n (-)^{l} w^{n-l} e_{l},
\label{eq:Qexp}
\end{equation}
where $e_{l}$ are the elementary symmetric functions (\ref{eq:esf})
with the variables $w_i$ as arguments.
Stroganov showed \cite{Str}, using Baxter's T-Q relation
\cite{Baxter} that $Q_n(w)$ can be calculated analytically in the case
where the $w_i$ are the solution of (\ref{eq:BAwPBC}) with $L=2n+1$
corresponding to the lowest value of the energy (\ref{eq:energy}). The
answer is given as a rational function in $w$. Using a binomial
coefficient identity, the explicit polynomial form of $Q_n(w)$ is
calculated in appendix \ref{ap:details}. Comparing the expression
(\ref{eq:QPBCpol}) now with the formal expansion of $Q_n(w)$ in
(\ref{eq:Qexp}) one can read off the values of the elementary
symmetric functions at these particular values of $w_i$. 
We find 
\begin{equation}
e_l = \binom{n-\frac{1}{3}}{n}^{-1}
\sum_{p=0}^{\lfloor l/3 \rfloor} \left[ \binom{2n-3p+l}{2n}
\binom{n-\frac{1}{3}}{n-p} \binom{n+\frac{1}{3}}{p}
- \binom{2n-3p+l-1}{2n} \binom{n-\frac{1}{3}}{p}
\binom{n+\frac{1}{3}}{n-p} \right]. 
\label{eq:elemPBC}
\end{equation}
The series in (\ref{eq:Qexp}) with the coefficients as in
(\ref{eq:elemPBC}) in general does not appear to be summable,
i.e. cannot be written as a simple product, but it can be verified
without too much difficulty that it satisfies the recursion relation
\begin{equation}
(w+1)^2 (3n+2)Q_{n+1}(w) = 3(w^3-1)(2n+1)Q_n(w) -
(w^2-w+1)^2(3n+1)Q_{n-1}(w).
\label{eq:recursPBC}
\end{equation}
For special values of $w$, $Q_n(w)$ simplifies dramatically.
The results 
\begin{equation}
q^{2n} Q_n(q^{-1})
= {2}^n \prod_{j=1}^n \frac{2j-1} {3j-1}, \quad Q_n(0) = (-)^n,
\label{eq:Qspec}
\end{equation}
can, e.g., be calculated easily from (\ref{eq:recursPBC}) when 
$w=q^{-1}$ and $w=0$.
{}From this we conclude that
\begin{equation}
\prod_{j=1}^n (1+z_j+z_j^2) = \left(\frac{3}{4} \right)^n
\prod_{j=1}^n \left(\frac{3j-1} {2j-1}\right)^2.
\end{equation}

Now let us rewrite the product
\begin{equation}
\prod_{i\neq j}^n (1+z_i +z_i z_j) = \prod_{i\neq j}^n \frac{\i
\sqrt{3} (q^2 w_j - w_i)} {(q w_i-1) (q w_j-1)} = \prod_{i=1}^n \left(
\frac{\sqrt{3}q^{-1}}{(w_i - q^{-1})^2} \right)^{n-1}
\prod_{i<j} \frac{w_i^3 -w_j^3}{w_i-w_j},  
\label{eq:prodPBC}    
\end{equation}
where in the last step we recognize the Schur function in the ratio of
the two Vandermonde determinants (see equation (\ref{eq:Schurdet})).
Thus we find, using (\ref{eq:Qspec}), 
\begin{equation}
\prod_{i\neq j}^n (1+z_i +z_i z_j) = \left(\prod_{j=1}^n
\frac{\sqrt{3}}{4} \left(\frac{3j-1}{2j-1}\right)^2 \right)^{n-1}
s_{(2(n-1),2(n-2),\ldots,2,0)}(w_1,\ldots,w_n).
\end{equation}
This can be rewritten using the fact that a Schur function
can be written as a determinant over the elementary symmetric
functions (see (\ref{eq:NK})).
We now have
\begin{equation}
s_{(2(n-1),2(n-2),\ldots,2,0)}(w_1,\ldots,w_n) =
\det \left( e_{n-[(i+1)/2]-i+j} \right)_{i,j}^{2(n-1)},
\label{eq:detPBC}
\end{equation}
with $e_l$ given by (\ref{eq:elemPBC}). When evaluated explicitly for
small values of $n$ we find 
\begin{equation}
\prod_{i\neq j}^n (1+z_i +z_i z_j) = A_n^3, \label{eq:conj}
\end{equation}
where $A_n$ is the number of $n\times n$ 
alternating-sign matrices (\ref{eq:asm}).
We have not been able to evaluate (\ref{eq:detPBC}) for arbitrary
values of $n$.

The smallest and largest component of the groundstate wave function
are given by $\psi(1,2,\ldots,n)$ and $\psi(1,3,\ldots,2n-1)$
respectively. In \cite{RS} it was conjectured that their ratio is
equal to $A_n$. In terms of the present approach, we find that the
smallest and largest component are given by
\begin{eqnarray} 
\sum_{\pi} \prod_{i < j} \frac{1+z_{\pi_i}^{-1} + z_{\pi_j}}{z_{\pi_j}
-z_{\pi_i}} &=& \sum_{\pi} \prod_{i < j} q^{-1} \frac{q w_{\pi_i}-1}{q
- w_{\pi_i}} \frac{w_{\pi_i} - q^2 w_{\pi_j}}{w_{\pi_j} -
w_{\pi_i}} = A_n, \\  
\sum_{\pi} \prod_{i < j} \frac{z_{\pi_i}^{-1} (1+z_{\pi_i}^{-1} +
z_{\pi_j}) }{z_{\pi_j} -z_{\pi_i}} &=& \sum_{\pi} \prod_{i < j} q^{-1}
\left(\frac{q w_{\pi_i}-1}{q - w_{\pi_i}} \right)^2 \frac{w_{\pi_i} - q^2
w_{\pi_j}}{w_{\pi_j} - w_{\pi_i}} = A_n^2. 
\end{eqnarray}
Since both functions above are (up to a common factor) symmetric
polynomials, we hope that these conjectures can be proven by making
use of (\ref{eq:elemPBC}).

\subsection{Twisted boundaries}

For twisted boundary conditions,
\begin{equation}
\sigma_{L+1}^{\pm} = (\sigma_{L+1}^x \pm \i \sigma_{L+1}^y) =
\e^{\pm2\i\phi} \sigma_{1}^{\pm},
\end{equation}
the eigenvectors and eigenvalues of $H$ are still given by
(\ref{eq:eigvec}) and (\ref{eq:energy}). The equations for $z_i$ or
$w_i$ however are modified and for the special case $\phi=\pi/3$ become 
\begin{equation}
\left( \frac{q-w_i}{q w_i-1}\right)^L + q^{-2}\prod_{j=1}^n
\frac{w_i - q w_j}{q w_i -w_j} =0. \label{eq:BAwTBC}
\end{equation}
Also for this case $Q_n(w)$ can be calculated analytically when the
solution of (\ref{eq:BAwTBC}) with $L=2n$ corresponds to the
groundstate \cite{FSZa}. We find that the elementary symmetric
functions now have the values 
\begin{equation}
e_l = \binom{n-\frac{1}{3}}{n}^{-1} \sum_{p=0}^{\lfloor l/3 \rfloor+1}
\left[ \binom{2n-3p+l-1}{2n-1} \binom{n-\frac{1}{3}}{n-p}
\binom{n-\frac{2}{3}}{p} 
 - \binom{2n-3p+l+1}{2n-1} \binom{n-\frac{1}{3}}{p-1}
\binom{n-\frac{2}{3}}{n-p} \right]. \label{eq:elemTBC}
\end{equation}
In a similar way as was done above for the periodic boundaries we can use
this expression to evaluate symmetric polynomials in the variables
$z_i$. For example, we find that
\begin{equation}
\prod_{i\neq j}^n (1+z_i +z_i z_j) = \left(4 \e^{-\i\pi/3} 3^{n/2-1}
\prod_{j=1}^n \left(\frac{3j-1}{n+j}\right)^2 \right)^{n-1} \det
\left( e_{n-[(i+1)/2]-i+j} \right)_{i,j}^{2(n-1)}. 
\end{equation}
When evaluated for small values of $n$ we see this is equal to
\begin{equation}
\prod_{i\neq j} (1+z_i+z_iz_j) = \e^{-(n-1)\i\pi/3} A_n
A_{\rm HT}(2n-1),
\label{eq:conj1}
\end{equation}
where
\begin{equation}
A_{\rm HT}(2n+1) = A_n^2 \prod_{k=1}^n \frac34 \left( \frac{3k-1}{2k-1} \right)^2
\end{equation}
is the number of $(2n+1) \times (2n+1)$ half-turn symmetric alternating 
sign matrices.

\subsection{Reflecting boundaries}

For the open chain with diagonal, or spin conserving, boundaries, the
Hamiltonian is given by
\begin{equation}
H = - \frac{1}{2} \sum_{j=1}^{L-1} \left( \sigma_j^x \sigma_{j+1}^x +
\sigma_j^y \sigma_{j+1}^y + \Delta \sigma_j^z \sigma_{j+1}^z -
\frac{1}{2}(q-q^{-1})(\sigma_1^z -
\sigma_L^z) \right). 
\label{eq:XXZhamOpen}
\end{equation}
The eigenvectors are now
\begin{equation}
\psi(x_1,\ldots, x_n) = \sum_{\pi,\sigma} A^{\pi_1\ldots
\pi_n}_{\sigma_1\ldots \sigma_n} \prod_{j=1}^n
z_{\pi_j}^{\sigma_j x_j},
\end{equation}
where the sum runs over all permutations $\pi = (\pi_1,\ldots,\pi_n)$
of the numbers $1,\ldots,n$ and all signs $\sigma_i = \pm 1$
\cite{ABBBQ,Gaudin}. In this case the energy is given by
\begin{equation}
E = -\frac{1}{2}(L-1)\Delta - \sum_{i=1}^n (z_i + z_i^{-1} -2\Delta).
\label{eq:energyOpen}
\end{equation}
The Bethe equations become
\begin{equation}
\left(\frac{q-w_i}{qw_i-1}\right)^{2L} - \prod_{
\stackrel{j=1}{\scriptscriptstyle j\neq i}}^n \left(\frac{q^2
w_j-w_i}{w_j - q^2 w_i}\right) \left(\frac{q^2 - w_i w_j}{1- q^2 w_i
w_j}\right) =0, 
\label{eq:BAOpen} 
\end{equation}
where $z$ and $w$ are again related by (\ref{eq:varchange}).
The amplitudes are up to a normalisation given by
\begin{equation}
A^{\pi_1\ldots \pi_n}_{\sigma_1\ldots \sigma_n} = \prod_{i=1}^n
z_{\pi_i}^{-\sigma_i L} \frac{1+q z_{\pi_i}^{-\sigma_i}}
{z_{\pi_i}^{\sigma_i} - z_{\pi_i}^{-\sigma_i}}
\prod_{i<l} \frac{(q^2 w_{\pi_i}^{-\sigma_i} - w_{\pi_l}^{-\sigma_l})
(q^2 - w_{\pi_i}^{\sigma_i} w_{\pi_l}^{\sigma_l})} {(w_{\pi_i}
-w_{\pi_l}) (1-w_{\pi_i}^{-1} w_{\pi_l}^{-1})}. 
\end{equation}
In this case the function $Q$ is defined by
\begin{equation}
Q_n(w) = \prod_{i=1}^n (\tilde{w}-\tilde{w}_i) = \sum_{i=0}^n (-)^i
\tilde{w}^{n-i} e_i (\tilde{w}_1, \ldots, \tilde{w}_n),\quad \tilde{w}
= w+w^{-1}.
\end{equation}
Also for this case $Q_n(w)$ can be given analytically when the solution
of (\ref{eq:BAOpen}) corresponds to the lowest value of the energy
(\ref{eq:energyOpen}) \cite{FSZa}.
The explicit polynomial form of $Q_n(w)$ is given in the Appendix, from
which the $e_i$ can be read off.  

In analogy with the periodic and twisted cases we consider the product
\begin{eqnarray}
\prod_{i=1}^{2n} \prod_{\stackrel{j=1}{\scriptscriptstyle z_j \neq
z_i^{\pm 1}}}^{2n} (1+z_i+z_iz_j) &=& \prod_{i=1}^{2n}
\prod_{\stackrel{j=1}{\scriptscriptstyle w_j \neq 
w_i^{\pm 1}}}^{2n} \frac{(q-q^{-1})(q^2 w_j-w_i)} {(qw_i-1)(qw_j-1)}
\nonumber\\
&=& \prod_{i=1}^n \left(\frac{q-q^{-1}} {\tilde{w}_i -q -q^{-1}}
\right)^{4(n-1)} \prod_{i\neq j}^n \frac{w_i^3 +w_i^{-3} - w_j^3 -
w_j^{-3}} {\tilde{w}_i - \tilde{w}_j} \nonumber\\
&=& \prod_{i=1}^n \left(\frac{q-q^{-1}} {\tilde{w}_i -q -q^{-1}}
\right)^{4(n-1)} \left( \frac{\det (w_i^{3(n-j)} + w_i^{-3(n-j)})}
{\det (w_i^{n-j} + w_i^{-n+j})} \right)^2.\label{eq:prodOBC}
\end{eqnarray}
Here we use the convention $w_{i+n} = w_i^{-1}$. The ratio of
determinants in (\ref{eq:prodOBC}) is up to a factor a symmetric polynomial in
$\tilde{w}_1 \ldots \tilde{w}_n$. Unfortunately we have not succeeded in
expressing it in the known basis of elementary symmetric functions. 
Nevertheless, numerical evaluation of (\ref{eq:prodOBC}) leads us 
to conjecture that
\begin{equation}
\prod_{i=1}^{2n} \prod_{\stackrel{j=1}{\scriptscriptstyle z_j \neq
z_i^{\pm 1}}}^{2n} (1 + z_i +z_i z_j) = A_{\rm V}(2n+1) ^2 N_8(2n)^4,
\label{eq:conj2}
\end{equation} 
where $A_{\rm V} (2n+1)$ is the number of $(2n+1) \times (2n+1)$
vertically symmetric alternating sign matrices (Ref.~\cite{K}, Theorem 2) given by
\begin{equation}
 A_{\rm V} (2n+1) = \prod_{j=0}^{n-1} (3j+2)
\frac{(2j+1)!(6j+3)!}{(4j+2)!(4j+3)!}
\end{equation}
and $N_8(2n)$ is the number of cyclically symmetric transpose
complement plane partitions \cite{Bress,RR} given by
\begin{equation}
N_8(2n) = \prod_{i=1}^{n-1} (3i+1) \frac{(2i)!(6i)!}{(4i)!(4i+1)!}.
\end{equation}

\section{CONCLUDING REMARKS}

We have made a first step towards proving the appearance of certain
numbers related to alternating sign matrices in the groundstate
eigenvector of the XXZ spin chain.
Many of the combinatorial results concerning alternating sign matrices
have been obtained using the connection with the integrable six-vertex
model \cite{K1,K}. The XXZ spin chain is closely related to the
six-vertex model and the methods used in this paper provide a
different application of integrability.  
The eigenvectors of the XXZ Hamiltonian are given in the form
of Bethe's Ansatz as a result of the integrability of the spin chain
for each of the different boundary conditions under consideration.
In fact, the normalisation of the amplitudes
(\ref{eq:amplitude}) ensures that all eigenvectors are symmetric 
polynomials in the Bethe roots. 
As is well known, a basis for the space of symmetric polynomials is
given by the elementary symmetric functions.
Using the results of Stroganov and his collaborators \cite{FSZa,FSZb,Str} 
for Baxter's $Q$-function, we were able to
derive explicitly the values that the elementary symmetric functions
take at the groundstate solution of the Bethe roots. 
See, e.g., equations (\ref{eq:elemPBC}) and (\ref{eq:elemTBC}) 
for the periodic and twisted cases.
Although in principle possible, we were not able to re-express the
groundstate in terms of the elementary symmetric functions.
However, we could show that certain natural double products over the 
Bethe roots can be rewritten in this way.
Using some results from the theory of symmetric functions we could 
then derive determinant expressions that
when evaluated give rise to the numbers counting
different symmetry classes of alternating sign matrices.
Our key results (\ref{eq:conj}), (\ref{eq:conj1}) and (\ref{eq:conj2})
thus remain as yet further conjectures.
It remains to be seen if such products over Bethe roots have any
direct combinatorial meaning.

\section{ACKNOWLEDGMENTS}
We thank Vladimir Mangazeev for stimulating discussions. This work has
been supported by The Australian Research Council.  

\appendix
\section{Details}
\label{ap:details}
\subsection{Periodic boundaries}

Stroganov's result for $Q_n(w)$ in the case of periodic boundary
conditions and odd system size is \cite{Str}
\begin{equation}
Q_n(w) =\prod_{i=1}^n (w-w_i)  
= \binom{n-\frac{1}{3}}{n}^{-1} \sum_{k=0}^n (-)^k
\binom{n-\frac{1}{3}}{k} \binom{n+\frac{1}{3}}{n-k}
w^{3k+1} \frac{(-1)^n + w^{3n-6k-1}}{(1+w)^{2n+1}}.
\label{eq:StrogQPBC} 
\end{equation}
We would like to rewrite this in the form
\begin{equation}
Q_n(w) = \sum_{i=0}^n (-)^{i} w^{n-i} e_{i},
\end{equation}
where the $e_{i}$ are the elementary symmetric functions with
arguments $w_1,\ldots,w_n$. The right hand side of (\ref{eq:StrogQPBC})
can be expanded using 
\begin{equation}
\sum_{m=0}^{\infty} (-)^m \binom{2n+m}{m} w^{-2n-m-1} =
(1+w)^{-2n-1}. \label{eq:sum1}
\end{equation}
It follows that $Q$ can be written as
\begin{eqnarray}
Q_n(w) &=&  \binom{n-\frac{1}{3}}{n}^{-1} \sum_{k=0}^n
\sum_{m=0}^{\infty} (-)^{k+m} w^{n-3k-m} \left[ \binom{2n+m}{m}
\binom{n-\frac{1}{3}}{n-k} \binom{n+\frac{1}{3}}{k}
\right. \nonumber\\ && \left. \hphantom{\binom{n-\frac{1}{3}}{n}^{-1}
\sum_{m=0}^{\infty} w^{n-3k-m} \left[ \right.} - \binom{2n+m-1}{m-1}
\binom{n-\frac{1}{3}}{k} \binom{n+\frac{1}{3}}{n-k} \right]. 
\label{eq:Qrew1}
\end{eqnarray} 
To show that (\ref{eq:Qrew1}) reduces to a finite sum, we first
collect terms of the form $m=3j$, $m=3j+1$ and $m=3j+2$. Then we use
the following identity to rewrite (\ref{eq:Qrew1}) 
\begin{equation}
\sum_{k=0}^n \sum_{j=0}^{\infty} a_{k,j} = \sum_{l=0}^{n}
\sum_{p=0}^{l} a_{l-p,p} + \sum_{l=n+1}^{\infty} \sum_{p=0}^{n}
a_{p,l-p}. \label{eq:resum}
\end{equation}
To proceed we need the result 
\begin{equation}
\sum_{p=0}^{n} \binom{3p-n+s}{2n} \binom{n-\frac{1}{3}}{p}
\binom{n+\frac{1}{3}}{n-p} = \sum_{p=0}^{n} \binom{3p-n+s-1}{2n}
\binom{n-\frac{1}{3}}{n-p} \binom{n+\frac{1}{3}}{p}.
\label{le:hyp1}
\end{equation}
This identity is proven by first showing that both the left and right
hand sides satisfy the same recursion relation in $n$, i.e. both sums
for $n=m+3$ can be expressed in the same sums for $n=m+2,m+1$ and
$n=m$. One then shows explicitly that the identity holds for
$n=0,1,2$. The recursion relation for (\ref{le:hyp1}) can be easily
derived using the Paule and Schorn Mathematica implementation of an
algorithm of Zeilberger's \cite{PS}.

{}From (\ref{le:hyp1}) it then follows that the infinite sum in
(\ref{eq:resum}) vanishes and after recollecting terms again we can
finally write   
\begin{eqnarray}
Q_n(w) &=& \binom{n-\frac{1}{3}}{n}^{-1} \sum_{l=0}^n
\sum_{p=0}^{[l/3]} (-)^l w^{n-l} \left[ \binom{2n-3p+l}{2n}
\binom{n-\frac{1}{3}}{n-p} \binom{n+\frac{1}{3}}{p}
\right. \nonumber\\ 
&& \left. \hphantom{\binom{n-\frac{1}{3}}{n}^{-1} \sum_{l=0}^n
\sum_{p=0}^{[l/3]} (-)^l w^{n-l} \left[ \right.} -
\binom{2n-3p+l-1}{2n} \binom{n-\frac{1}{3}}{p}
\binom{n+\frac{1}{3}}{n-p} \right], 
\label{eq:QPBCpol}
\end{eqnarray}
the desired result.

\subsection{Twisted Boundaries}

In this case the result of Fridkin et al. \cite{FSZa} for $Q_n(w)$ is
\begin{equation}
Q_n(w) = \binom{n- \frac{1}{3}}{n}^{-1} \frac{1}{(1+w)^{2n}}
\sum_{k=0}^n (-)^k \binom{n-\frac{2}{3}}{n-k} \left( (-)^n
\binom{n-\frac{1}{3}}{k} w^{3k} - \binom{n-\frac{1}{3}}{k-1}
w^{3n-3k+2} \right). 
\end{equation}
In analogy with the periodic case we need to rewrite $Q_n(w)$ in powers of
$w$. This can be done along the lines of the previous subsection with
the help of the result
\begin{equation}
\sum_{p=0}^{n} \binom{3p-n+s}{2n} \binom{n-\frac{1}{3}}{p}
\binom{n-\frac{2}{3}}{n-p} = \sum_{p=0}^{n} \binom{3p-n+s+2}{2n}
\binom{n-\frac{1}{3}}{n-p-1} \binom{n-\frac{2}{3}}{p}.
\label{le:hyp2}
\end{equation}
As in the case of (\ref{le:hyp1}), this identity can be proven by
showing that both the left and right hand sides satisfy the same
recursion relation in $n$. 
We then find 
\begin{eqnarray}
Q_n(w) &=& \binom{n-\frac{1}{3}}{n}^{-1} \sum_{i=0}^n (-)^i w^{n-i}
\sum_{p=0}^{\lfloor l/3 \rfloor+1} \left[ \binom{2n-3p+l-1}{2n-1}
\binom{n-\frac{1}{3}}{n-p} \binom{n-\frac{2}{3}}{p} \right. \nonumber\\
&& \left. \hphantom{c_0 \sum_{l=0^n} \sum_{p=0}^{l/3} \left[
\right.} - \binom{2n-3p+l+1}{2n-1} \binom{n-\frac{1}{3}}{p-1}
\binom{n-\frac{2}{3}}{n-p} \right]. \label{eq:QTBC}
\end{eqnarray}
\subsection{Reflecting boundaries}

Following Fridkin et al. \cite{FSZa} one can prove that $Q$ is given by
\begin{equation}
\binom{2n-\frac{2}{3}}{2n}^{-1} \sum_{k=-n}^n
(-)^{n+k}\binom{2n+\frac{2}{3}}{n-k} \binom{2n-\frac{2}{3}}{n+k}
\frac{w^{3k+1} - w^{-3k-1}}{(w-w^{-1})(2+w+w^{-1})^{2n}}.
\end{equation}
Using (\ref{eq:sum1}) and
\begin{equation}
\sum_{p=0}^{\lfloor n/2\rfloor} (-)^p \binom{n-p}{p} (w+w^{-1})^{n-2p} =
\frac{w^{n+1} - w^{-n-1}} {w-w^{-1}}
\end{equation}
the expression for $Q$ can be rewritten so that we can read off the
values of the elementary symmetric functions with arguments
$\tilde{w}_1, \ldots, \tilde{w}_n$, where $\tilde{w}=w+w^{-1}$. 
Namely 
\begin{eqnarray}
Q_n(w) &=& 2^{-3n} \binom{2n-\frac{2}{3}}{2n}^{-1} \sum_{p=0}^n (-2)^p
\tilde{w}^{n-p} \left[ \binom{3n-p-1}{n-p}\binom{2n+\frac{2}{3}}{n}
\binom{2n+\frac{2}{3}}{n} + {} \right. \nonumber\\ 
&& {}\sum_{m=0}^p \sum_{k=1}^n
(-)^{[(k+m+1)/2]} 2^{m-1} \left( 1 + (-)^{k+m} \right)
\binom{3n-p-m-1}{2n-1} \times \nonumber\\ 
&& \left.\left( \binom{2n+\frac{2}{3}}{n-k}
\binom{2n+\frac{2}{3}}{n+k} \binom{\left[\frac{3k+m+1}{2}\right]}{m}
+ \binom{2n+\frac{2}{3}}{n+k} \binom{2n+\frac{2}{3}}{n-k}
\binom{\left[\frac{3k+m-1}{2}\right]}{m}\right)\right]. \nonumber\\ 
\end{eqnarray}

\end{document}